\documentclass{aastex631}

\newcommand{\WF}{{\em Roman}}

\newcommand{\wband}{$W146$-band}
\newcommand{\zband}{$Z087$-band}

\begin{document}

\title{The Scientific Discovery Space for the \WF\, Galactic Bulge Time Domain Survey}

\author{Jennifer C. Yee}
\affiliation{Center for Astrophysics $|$ Harvard \& Smithsonian, 60 Garden
St., Cambridge, MA 02138, USA}

\author{Andrew Gould}
\affiliation{Max-Planck-Institute for Astronomy, K\"{o}nigstuhl 17,
69117 Heidelberg, Germany}
\affiliation{Department of Astronomy, Ohio State University, 140 W.
18th Ave., Columbus, OH 43210, USA}

\section{State of the Field}

\begin{figure}
	\hspace*{0.85cm}\includegraphics[width=0.9\textwidth]{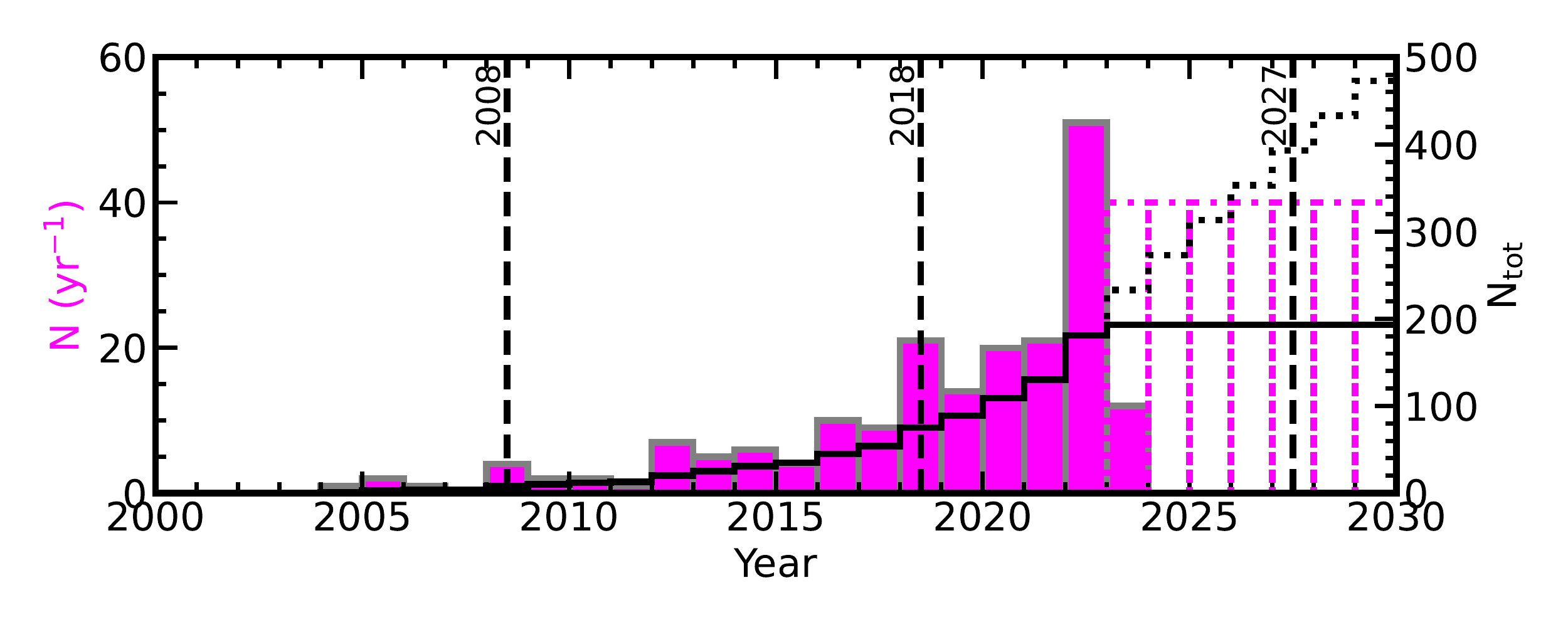}
	\includegraphics[width=0.9\textwidth]{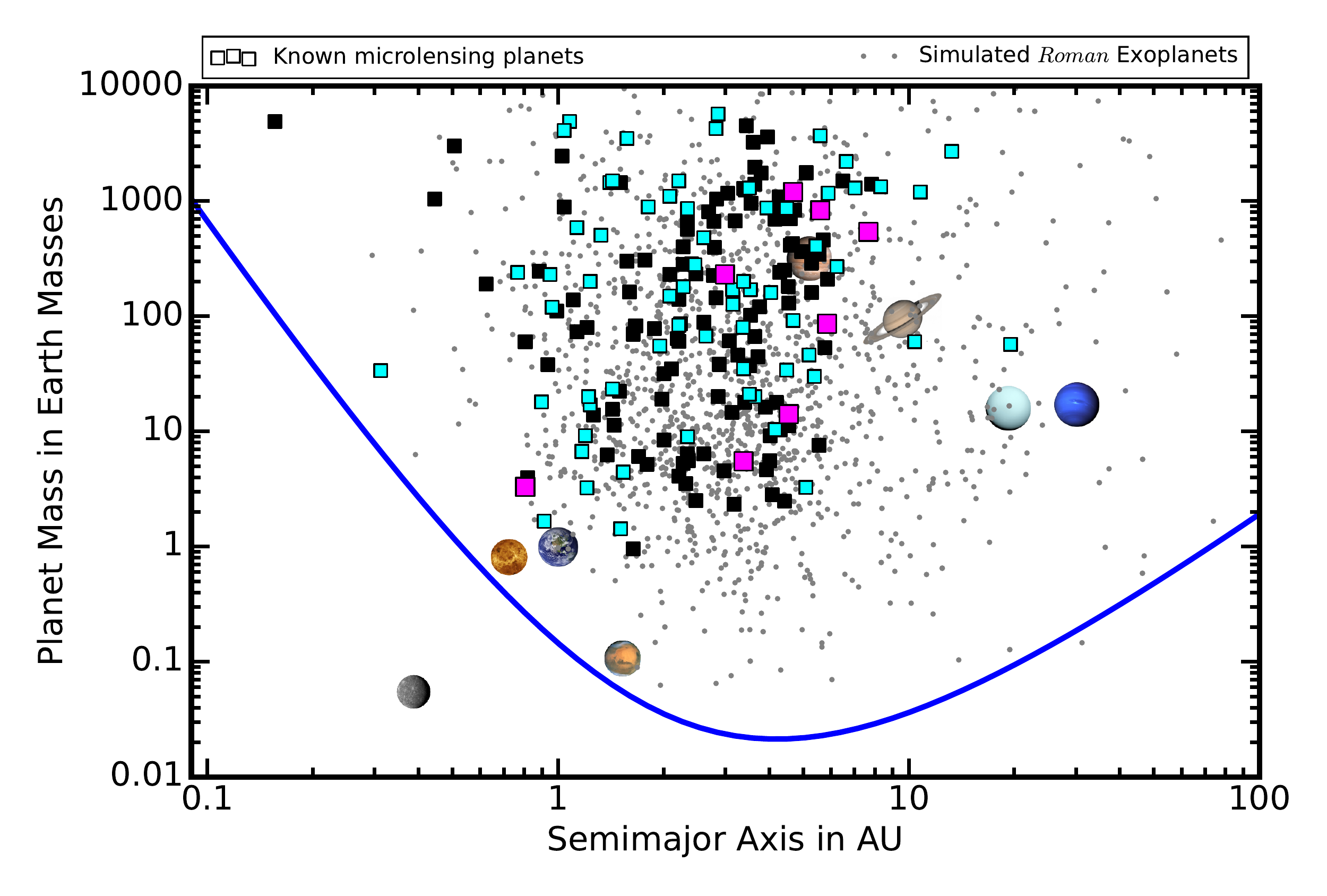}
	\caption{\WF\, Galactic Bulge Time Domain Survey discovery space compared to microlensing discoveries over time (NASA Exoplanet Archive, accessed 5/19/23). Top, left axis (magenta): number of microlensing planets published each year. A publication rate of 40 planets/year from the ground is assumed for future years. Top, right axis (black): cumulative number of published microlensing planets (solid=actual, dotted=projected). The black dashed lines indicate the approximate times when white papers were being prepared for the 2010 and 2020 decadal surveys and the start of the \WF\, Galactic  Bulge Time Domain Survey. Bottom \citep{Penny19}: squares = known planet distribution at the end of 2008 (magenta), 2018 (cyan), and now (black) compared to the Solar System planets and simulated \WF\, planet yield (gray points). The blue line indicates the expected \WF\,, sensitivity limit.\label{fig:discovery_space}}
\end{figure}

The microlensing planet landscape has changed dramatically since a space-based microlensing mission was recommended by the 2010 Decadal Survey. At the end of 2008, when white papers were being prepared for the decadal, only eight microlensing planets had been detected from ground-based surveys (Figure \ref{fig:discovery_space}). In that context, a space-based mission that could detect hundreds to thousands of planets was an obvious advance. Even toward the end of 2018, when almost 100 microlensing planets had been detected, there were only a handful of non-giant planet detections. However, today, the total number of planets has doubled and is projected\footnote{KMTNet currently detects about 30 planets per year but is also still dealing with a publication backlog, so we have assumed a future publication rate of 40 planets per year.} to double again prior to the start of the \WF\, Galactic Bulge Time Domain Survey (RGBS). In addition, there are now detections from the ground of several dozen planets with masses between 1 and 10 $M_\oplus$, a number that should also roughly double over the next few years. Hence, designing an optimal RGBS requires carefully considering the new scientific discovery space opened up by \WF.

Comparing the known distribution of microlensing planets (NASA Exoplanet Archive, accessed 5/19/23) and the simulated \WF\, detections from \citet{Penny19} in Figure \ref{fig:discovery_space} readily reveals two of the major strengths of RGBS compared to what can be (and has been) done from the ground. First, RGBS is projected to find planets in wider orbits, out to and beyond the orbits of Uranus and Neptune. Second, RGBS should find planets as small as or smaller than Mars. In addition, RGBS will be better than ground-based surveys at characterizing the smallest free-floating planets \citep{Gould23_kb2397}. All of these discovery spaces share the characteristics of having a low-cross section for detection, short duration signals, and potentially lower amplitude signals, and so lead to similar considerations for enhancing their detection rates.

In this white paper, we will begin by reviewing the relative sensitivity of RGBS relative to existing ground-based surveys. We will then discuss the challenges of each of the three discovery spaces described above and how they affect survey design. Because these challenges are mostly rooted in light curve degeneracies, these sections are necessarily somewhat technical. We conclude with a high-level summary including our recommendations.

\vspace{12pt}
{\section{A Change of Parameters}}

\begin{figure}
\begin{centering}
	\includegraphics[width=0.9\textwidth]{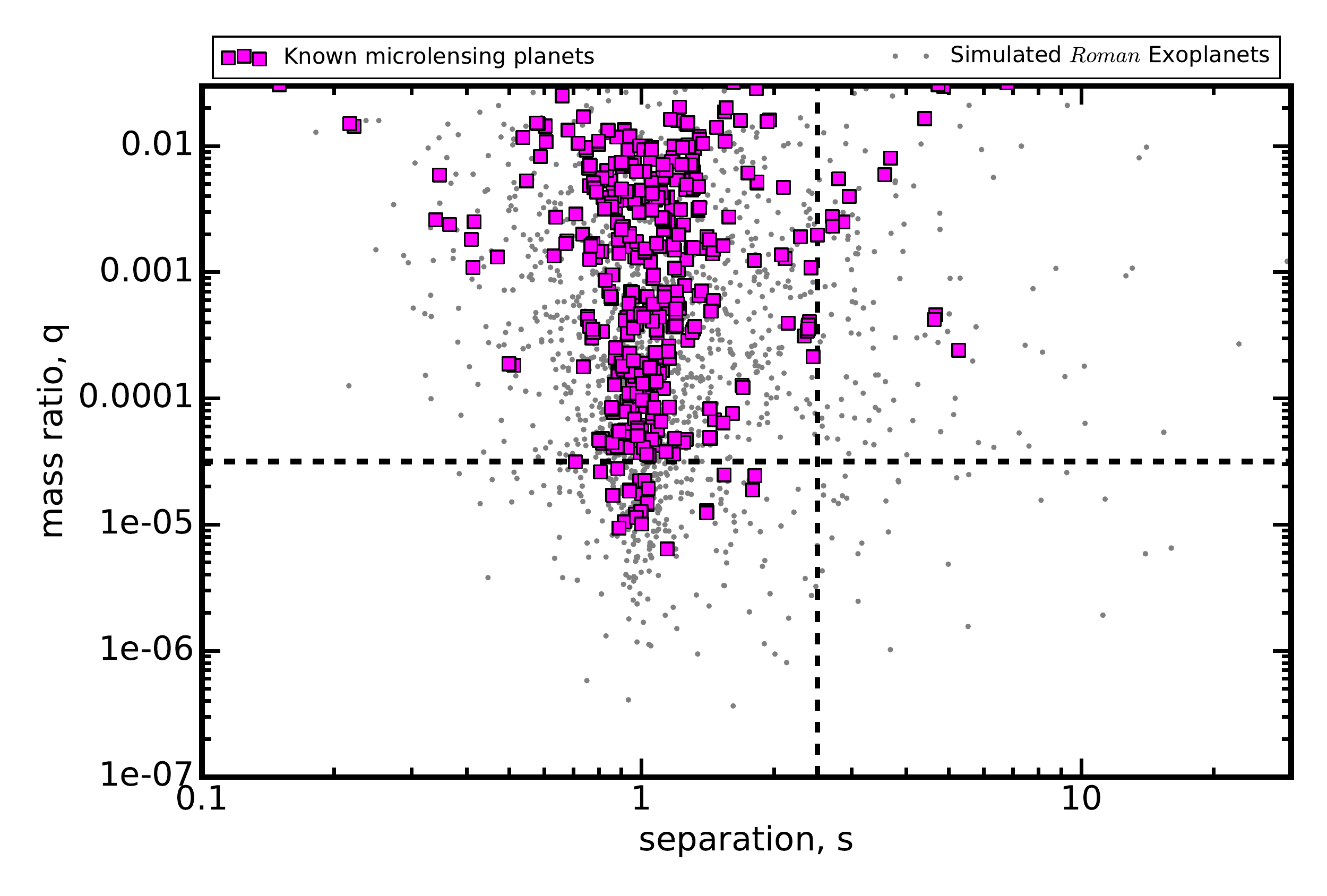}
	\caption{Known microlensing planets (magenta) compared to simulated RGBS planets (gray) in the microlensing parameter space: $s$, $q$. The new discovery space opened up by \WF\, will be for wide-orbit ($s \gtrsim 2.5$) and small ($q \lesssim 3\times 10^{-5}$) planets (black dotted lines).\label{fig:sq}}
\end{centering}
\end{figure}

Figure \ref{fig:discovery_space} shows planet mass vs. semi-major axis. However, the actual microlensing observables are planet mass ratio and projected separation scaled to the Einstein radius:
\begin{equation}
q \equiv \frac{m_{\rm p}}{M_{\rm host}}; 
\quad 
s \equiv \frac{a_{\perp}}{\theta_{\rm E} D_{\rm host}}, 
\end{equation} respectively, where
\begin{equation}
\theta_{\rm E} = \sqrt{\kappa M_{\rm host}{\pi_{\rm rel}}} 
\quad \mathrm{and} \quad
\pi_{\rm rel} = \mathrm{AU} \left(\frac{1}{D_{\rm host}} - \frac{1}{D_{\rm source}}\right).
\end{equation}
Therefore, Figure \ref{fig:discovery_space} is smeared out over the distribution of host masses and distances as well as the distances to the sources.

To illuminate the underlying the sensitivity space for RGBS, we re-plot the planet discoveries in Figure \ref{fig:sq} using the parameters $s$ and $q$. For simplicity, for planet detections, we use the default solution from the Exoplanet Archive, but as we will discuss below, the degeneracies can be quite severe.
The basic regions of interest remain the same, but we now give them formal definitions. For the purposes of this white paper, we define wide-orbit planets as those with $\log s > 0.4$ and small planets as those with $\log q < -4.5$. We will define ``free-floating planets" as any short timescale ($\lesssim 1\, $ day) signal without a microlensing light curve signature from the host.\footnote{Based on the light curve alone, these are necessarily only candidates because a host may simply be too distant to create a significant lensing signal, although the existence or absence of a host star can be further constrained by \WF\, limits on host light.}

\vspace{12pt}
{\section{\WF\, vs. KMTNet}
\label{sec:ag_comp}}

\begin{figure}
\begin{centering}
	\includegraphics[width=0.5\textwidth]{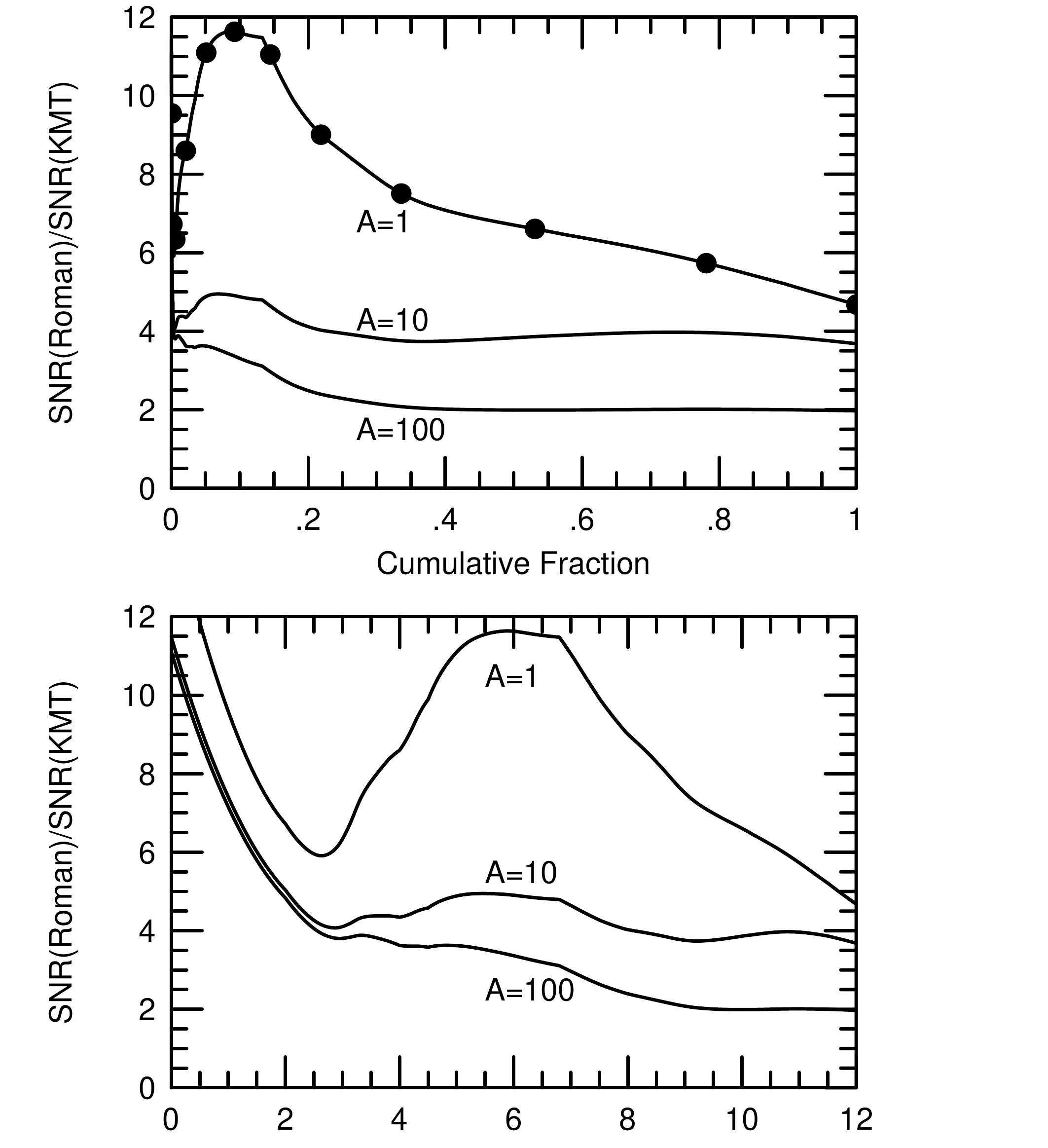}
	\includegraphics[width=0.5\textwidth]{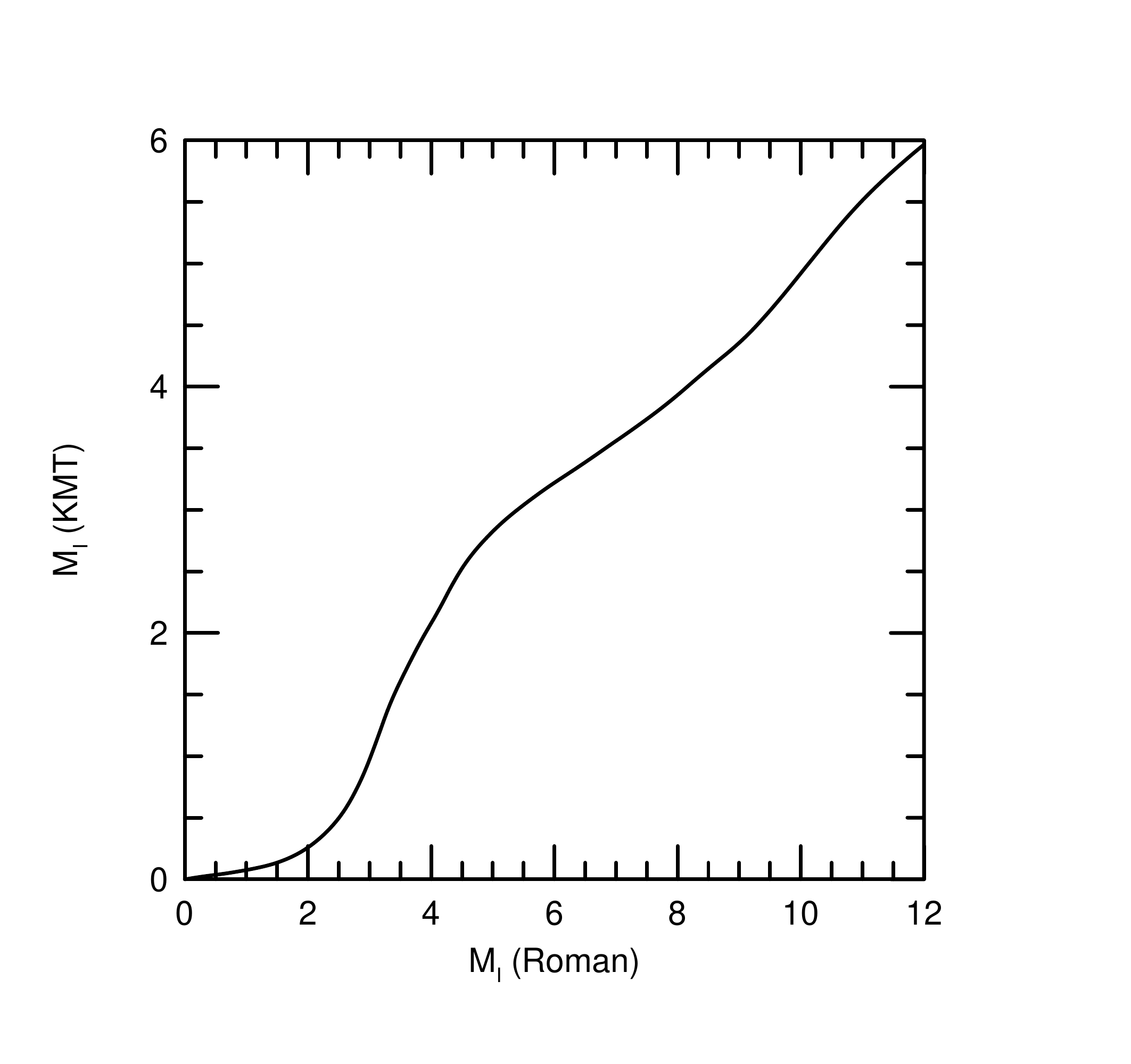}
	\caption{\WF's primary advantage over ground-based surveys comes at low magnification. For $A = 1$, \WF\, has a signal-to-noise advantage of $\sim 5$--$12$ over the KMTNet, but at $A=100$ that advantage is a relatively modest factor of $\sim 2$.
	Bottom: Absolute magnitudes of RGBS sources compared to KMTNet sources at the same cumulative fraction; e.g., half of all RGBS sources will be $M_I < 6$ whereas half of all KMTNet sources are $M_I < 3.25$. Top: relative signal-to-noise ratio between \WF\, and KMTNet as a function of cumulative fraction of sources and magnification. Middle: same as top, but as a function of \WF\, absolute source magnitude, for which the corresponding KMTNet source magnitude can be found from the bottom panel; e.g., for a \WF\, source with $M_I = 6$, the corresponding KMTNet source will be $M_I = 3.25$, and for such sources, \WF\, has a SNR advantage of $\sim 11.5$ over KMTNet at $A=1$, but only $\sim 3$ at $A=100$. Figures from \citet{Gould23_kb2397}.\label{fig:wf_kmt}}
\end{centering}
\end{figure}

The qualitative comparison between RGBS and the ground shown in Figure \ref{fig:discovery_space} neglects some potentially important details. In particular, the ground-based planets are actual detections whereas the RGBS planets are simulated by extrapolating an assumed mass function into new regimes. The mass function used in the simulations shown \citep{Penny19} assumes the occurrence rate of planets is flat (in log space) below $m_{\rm p} = 5.2 M_{\oplus}$, whereas \citet{Suzuki16} found a sharp break in the mass-ratio distribution with few or no planets smaller than $\log q \sim -5$. If the true mass function has fewer smaller or wide-orbit planets than assumed, the qualitative differences between RGBS and the ground would likely be reduced. Fortunately, \citet{Gould23_kb2397} have shown that RGBS should, in fact, be intrinsically more sensitive to planet detections than the ground and also that that advantage is most pronounced at low magnifications, e.g., for wide-orbit or free-floating planets.

In brief, \citet{Gould23_kb2397} showed that 10 years of the Korea Microlensing Telescope Network (KMTNet) survey will yield of order 10 times as many microlensing events as RGBS, but that RGBS has an advantage in signal-to-noise ratio. This advantage in signal-to-noise ratio arises from the difference in the sources probed by each survey and RGBS's lower background (mainly due to lower sky brightness).

Because RGBS is deeper than KMTNet, the typical source in RGBS is intrinsically fainter and smaller. The bottom panel of Figure \ref{fig:wf_kmt} shows the result of matching the cumulative fractions of KMTNet and RGBS sources to quantify this effect; i.e., a given fraction of RGBS sources will all be brighter than some magnitude, $M_I$ (\WF) and for the same fraction, KMTNet sources will all be brighter than some (brighter) magnitude, $M_I$ (KMT). Then, any given RGBS source can be matched to the corresponding KMTNet source, and the total signal-to-noise ratios for those sources can be directly compared (middle and top panels of Figure \ref{fig:wf_kmt}). At high magnification, $A$, events tend to be above the background in both KMTNet and RGBS, but at low magnification, e.g. $A\sim1$, RGBS sources remain above the background whereas KMTNet sources are below. This gives RGBS a large advantage in SNR at low magnifications.

Low magnifications also correspond to the regimes of interest for wide-orbit, small, and free-floating planets. Because wide-orbit planets produce planetary caustics far from their host stars, if the planets are detected through that channel (rather than from central caustic perturbations), the perturbations must occur at low magnification ($A\sim 1$). The situation is analogous for free-floating planets. For small planets, ground-based sensitivity is limited to detections with $\log s \sim 0$, for which the caustic structures occur at moderate to high magnification \citep{Abe13, Yee21_ob0960}. By contrast, to the extent that \WF\, detections are coming from planetary caustic perturbations, those will also tend to be at lower magnification. 

In conclusion, the work of \citet{Gould23_kb2397} shows that regardless of the intrinsic mass (or mass-ratio) distribution, RGBS has the most advantage over the ground for wide-orbit, small, or free-floating planet detections.

\vspace{12pt}
\section{Detection \& Characterization}

\subsection{Wide-Orbit Planets}

\begin{figure}
	\begin{centering}
	\includegraphics[height=0.7\textheight]{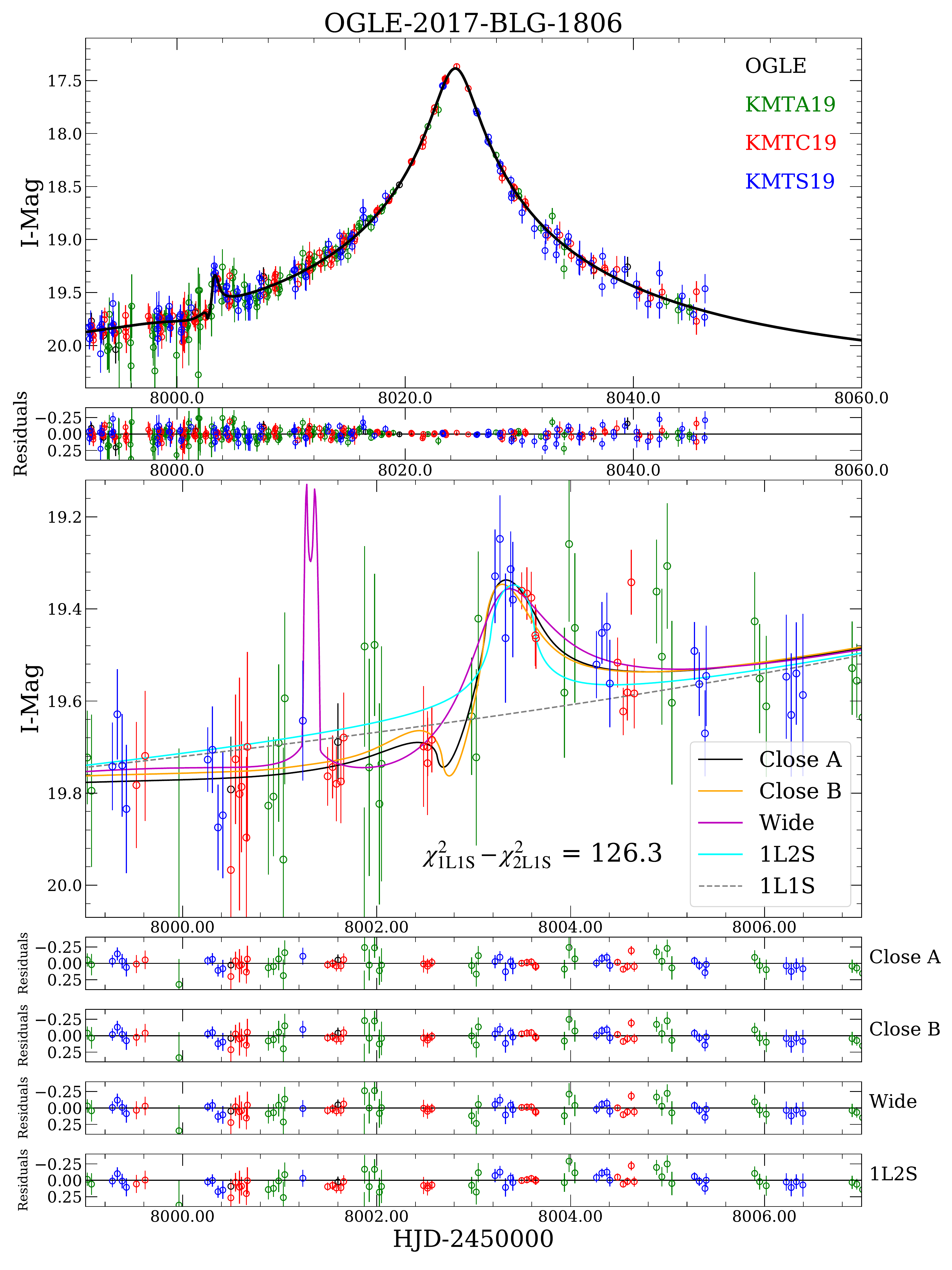}
	\caption{The ``bump"-like anomaly in OGLE-2017-BLG-1806 has multiple possible solutions with different values of the separation $s$ and mass ratio $q$. Such anomalies must also be checked for the 1L2S (binary source) degeneracy, which has a second source star instead of a planet \citep[Figure from ][]{Zang23AF7}.\label{fig:ob1806}}
	\end{centering}
\end{figure}

The scientific impact of RGBS depends not only on simply detecting wide-orbit planets, but also on the ability of RGBS to characterize and interpret them. If these planets can be successfully characterized, RGBS will open a new region of parameter space that will enable a measurement of the planet occurrence rate as a function of separation from $\lesssim 1\, $ au to $\gtrsim 30$ au.

The region of planet-host separation that is of greatest interest for RGBS is planets with $\log s > 0.4$ (i.e, $s \gtrsim 2.5$). 
In this regime, the caustics are well-separated, so a planet may either be detected through a central caustic perturbation  or a planetary caustic perturbation. Both of these regimes are subject to degeneracies.

Central caustic detections very often suffer from an intrinsic $s \rightarrow s^{-1}$ degeneracy \citep{GriestSafizadeh98} meaning that they cannot be unambiguously identified as wide-orbit planets.
While they are rare relative to the total number of {\it events}, they may contribute a larger fraction of detected {\it planets} due to the fact that the planet perturbations occur near the peak of the event. This makes them relatively easier to detect because the anomalies occur when the events are  brighter and the underlying stellar events are easier to detect. However, if they suffer from a severe degeneracy in $s$, they are not good candidates for a study of the separation distribution of planets. Hence, it is crucial to investigate what fraction of \WF\, wide-orbit planets detections will be due to central caustics, and among these, for what fraction (if any) the $s \rightarrow s^{-1}$ degeneracy can be resolved.

While planetary caustic perturbations do not suffer from the $s \rightarrow s^{-1}$ degeneracy, they are subject to a number of other degeneracies that may lead to ambiguity in interpreting an observed anomaly. Such planet perturbations manifest as ``bump-type" anomalies, i.e., the planet causes an increase in source magnification. 
Bump-like anomalies may also be produced by either a binary source \citep[1L2S; ][]{Gaudi98} or interacting with the triangular caustic of a larger $q$ planet with $s < 1$. In some cases, such anomalies may also suffer from the von Schlieffen vs. Cannae degeneracy, which can lead to ambiguities in both mass ratio $q$ and the source size $\rho$, affecting the physical interpretation of the event \citep{Hwang18_0173}.
Finally, there can also be degeneracies as to which side of the caustic the source passes, i.e., the ``inner-outer" degeneracy \citep{GaudiGould97}. Even if unresolved, the uncertainties due to the ``inner-outer" degeneracy are usually without major physical implications, so we do not discuss them here. 

For illustration, Figure \ref{fig:ob1806} shows an example of some of these potential degeneracies as seen in ground-based data. In this event, we see that there are subtle and not so subtle differences in the light curves of the various models. Hence, light curves with sufficient cadence and precision should be able to resolve these various degeneracies based on the detailed morphology of the perturbation. However, this is just one example, and in this example, the source crosses the caustic structure leading to a distinct, localized, perturbation. By contrast, \citet{Zhu14} predicted that about half of all planets would be detected from non-caustic crossing perturbations, and \citet{Jung23AF8} has confirmed this prediction for KMTNet. These perturbations can be much weaker making the potential degeneracies more severe.

A detailed investigation is needed to explore how these degeneracies vary with various properties, such as the amount of blending, the source size, and the location of the source trajectory relative to the caustic (which affects the strength of the perturbation). This simulation will need to determine the cadence of observations needed to resolve these degeneracies given the expected photometric precision of \WF\, and also the fraction of cases for which the degeneracies cannot be resolved.

An additional consideration for planetary caustic perturbations is that they are likely to occur well away from the peak of the stellar event.
For example, with $\log s = 0.4$ the perturbation will occur at $\tau \equiv (t_0 - t) / t_{\rm E} \gtrsim 2.1$. Hence, for a typical event with $t_{\rm E} = 20\, $ days, the interesting wide-orbit planets will all have $\Delta t > 42\, $ days. So for many of them, the peak of the event will not be captured due to the short ($\sim 72\, $ day) \WF\, observing season. 

This poses two fundamental challenges. First, it may be more challenging to detect the event in the first place due to the lack of signal from the stellar event. Second, there may be fundamental challenges to characterizing the planet. When the peak of an event is not captured, there is uncertainty in $u_0$ which leads to ambiguities in $t_{\rm E}$. These uncertainties can then propagate to uncertainties in $s$ and $q$. Furthermore, such partial light curves may also lead to uncertainties in the measurement of the microlens parallax, one of the observables used to constrain the masses of the host star. The potential for uncertainty in these parameters for wide-orbit planets should be investigated in more detail.

However, this degeneracy can also be mitigated through additional observations. The problem of characterizing the underlying stellar event will not be resolved by \WF\, observations in subsequent observing windows ($\sim 110\, $ days later), because the events will already have returned to baseline. But it can be mitigated by either increasing the duration of the \WF\, observing window or obtaining observations from another observatory (e.g., the ground) during the RGBS gaps to capture and characterize the peak of the light curve. The latter would be facilitated by choosing lower extinction fields that are accessible with ground-based optical telescopes.

\subsection{Small Planets}

\begin{figure}
	\begin{centering}
	\includegraphics[height=0.7\textheight]{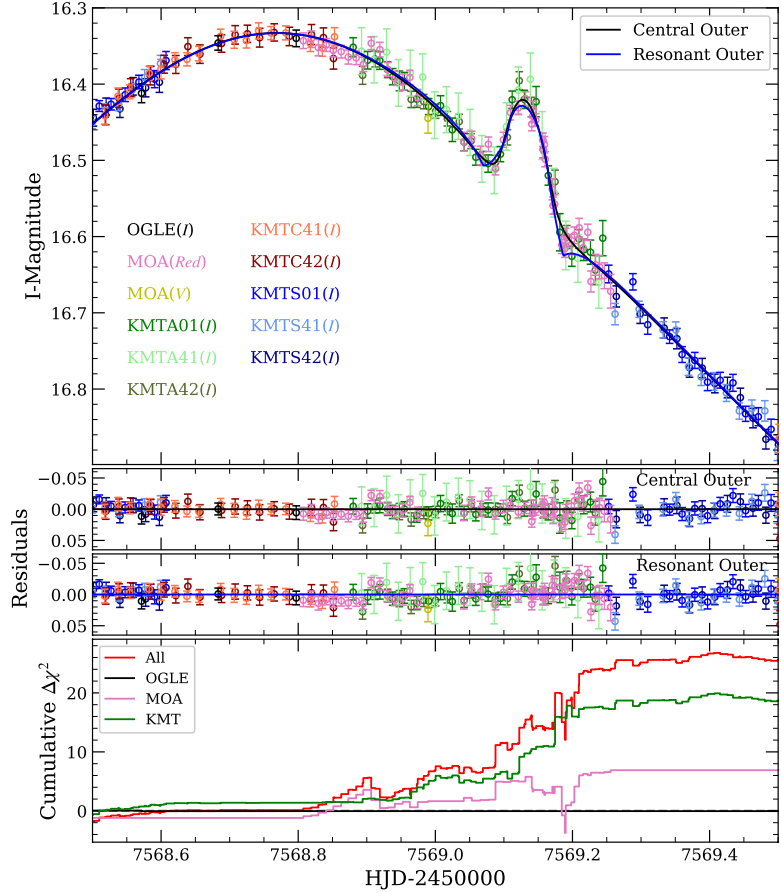}
	\caption{Small mass ratio planets can produce ``bump"-like perturbations that may be subject to degeneracies such as the ``central-resonant" degeneracy. This example shows two possible models for the planet in OGLE-2016-BLG-1195. Similar to potential degeneracies for wide-orbit planets, the RGBS cadence will need to be high enough to distinguish between various possible models. \citep[Figure from ][]{Gould23_ob1195} \label{fig:ob1195}}
	\end{centering}
\end{figure}

The issues for small ($\log q \lesssim -4.5$) are similar to the issues for wide-orbit planets. Some fraction of those planets will be detected as short-duration, ``bump"-like anomalies. As an example, consider Figure \ref{fig:ob1195}, which shows a ``bump"-like perturbation due to a $\log q \sim -4.5$ planet \citep{Gould23_ob1195}. This perturbation can be roughly described by either a central or a resonant caustic model. In these two different models, $q$ differs by a factor of $\sim 2$ and there is also a 10\% difference in $\rho$, which will affect the measurement of the Einstein radius, $\theta_{\rm E}$, and so the constraint on the host mass. There are also other cases in which this type of degeneracy leads to larger uncertainties in $\rho$ \citep[e.g.,][]{Ryu22_MP1}. Binary source solutions must also be considered for the case shown in Figure \ref{fig:ob1195}. 

So, the question is the same as for wide-orbit planets: Given the expected photometric precision, is the cadence high enough to measure the detailed light curve morphology with sufficient accuracy to distinguish between potentially degenerate solutions? 

In addition, the perturbation in Figure \ref{fig:ob1195} also suffers from the $s \rightarrow s^{-1}$ degeneracy. Because this degeneracy is mathematical in nature, it is often unresolved. Although it does not affect our ability to measure the mass ratio function, we should consider the potential impact on our ability to measure the separation distribution of small planets. For example, if (as we very roughly estimate) 2/3  of detectable planetary anomalies suffer 
from this degeneracy, we would need three times
 as many planets to achieve a given statistical precision, and so would need to monitor three times as many stars.

\subsection{Free-Floating Planets}

\begin{figure}
\begin{centering}
	\includegraphics[width=0.6\textwidth]{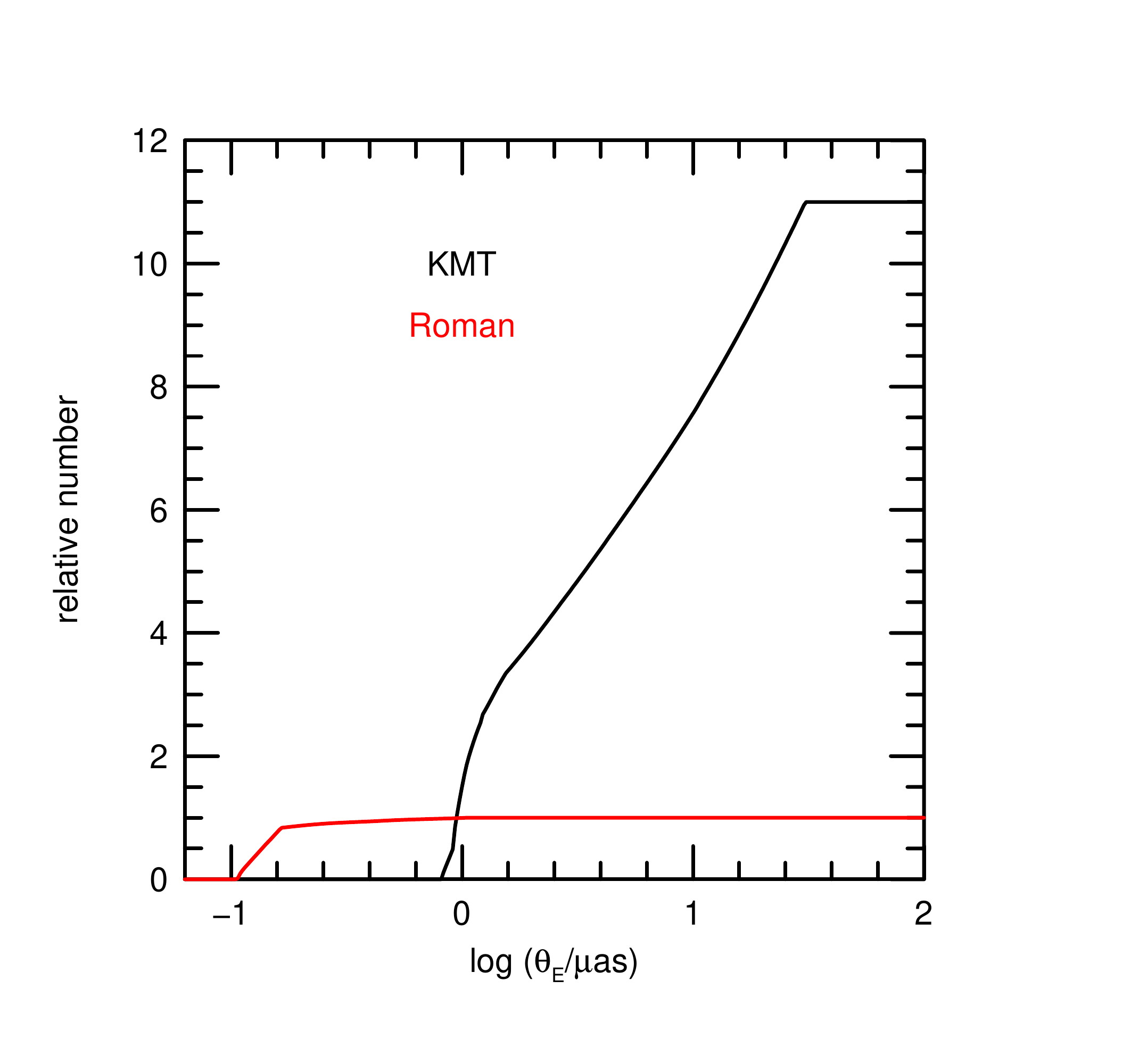}
	\caption{Relative expected number of finite-source point-lens events detected by a 10-yr KMT survey compared to \WF\, as a function of the Einstein radius, $\theta_{\rm E}$, which is proportional to the square-root of the lens mass. \WF\, has the most scientific power for $\theta_{\rm E} < 1 \mu{\rm as}$, so the characterization of those events is what will set the requirements for the RGBS cadence. Figure from \citet{Gould23_kb2397}\label{fig:fspl}}
\end{centering}
\end{figure}

The majority of RGBS free-floating planet candidates are essentially an extreme limit of wide-orbit planetary-caustic detections. Once they have been detected, the major issue for characterizing them is assessing whether or not they have a host star. One aspect of this is searching for an underlying stellar microlensing event, which might peak outside the \WF\, observing window. As with wide-orbit planets this search would benefit from observations between \WF\, observing windows from another observatory. Another aspect of host searches would be to search for a perturbation in the free-floating planet candidate event due to the host star's effect on the planetary caustic. Again, as with wide-orbit planets, this requires that the cadence to be high enough to measure the detailed light curve morphology, although tests need to be done to establish whether the required cadence is different from wide-orbit planets with distinct stellar events.

For the minority of RGBS events with measurable finite-source effects, \citet{Gould23_kb2397} have shown that \WF\, has a substantial advantage over KMTNet only for the smallest Einstein radii: $\theta_{\rm E} / \mu{\rm as} \lesssim 1$. See Figure \ref{fig:fspl}. Such events are extremely valuable because finite-source effects allow a measurement of $\theta_{\rm E}$, which is a much stronger constraint on the lens mass than $t_{\rm E}$ alone. Furthermore, the small $\theta_{\rm E}$ regime that RGBS will dominate also corresponds to the smallest free-floating planet candidates ($\sim 1$ to a few $M_\oplus$). This is the regime that RGBS can contribute most to measuring the mass function of free-floating planet candidates. Hence, the cadence of \WF\, needs to be high enough to measure the finite-source effect, i.e., $\rho$, for this class of event.

\vspace{12pt}
\newpage
\section{A Note on Colors/Multi-band Observations}

The nominal RGBS will take most of its observations in \wband, with observations in \zband\, every 12 hours. In principle, such observations can be used to confirm the microlensing nature of a perturbation (which should be achromatic). In practice, this can almost always be determined on the basis of light curve morphology. So, the primary purpose of the \zband\, observations is to measure the colors of the sources. 

In principle, if a given perturbation is due to a binary source rather than a planet, and if the two components of the binary source have significantly different colors, it is possible to rule out a planetary solution if a change in color is observed during the perturbation. However, if the color remains the same, depending on the flux ratio between the sources, a binary source solution is not necessarily ruled out because the two source stars may simply have the same color. 
In addition, in practice, the short durations of the perturbations mean that it is relatively unlikely that a \zband\, observation will be taken during the perturbation, so it will be difficult to measure such a color change. 

Due to limitations in the number of filter wheel changes for \WF, it is doubtful that the cadence of \zband\, observations can be increased sufficiently to guarantee a color measurement during the short perturbations that are typical of wide-orbit or small planets. This could be investigated in more detail, which would require assessing both the fraction of cases for which a significant color change would be expected as well as the number of \zband\, observations required to measure it. However, because at least 2 or 3 \zband\, observations would likely be required during a few hour perturbation,
color measurements are not likely to be a viable channel for distinguishing between binary source and planetary models.
Hence, the cadence of \zband\, observations only needs to be sufficient to measure the color of the primary source, which may imply that the cadence can be further reduced from the nominal $1/12\, {\rm hr}^{-1}$.

\section{Summary}

Maximizing the scientific return of \WF\, requires focusing on the scientific discovery space opened up by \WF\, relative to the ground: i.e., planets in wide orbits ($\log s \gtrsim 0.4$), the smallest mass-ratio planets ($\log q \lesssim -4.5$), and free-floating planet candidates (especially those with $\theta_{\rm E} \lesssim 1 \mu{\rm as}$). However, capitalizing on that leverage requires not just detecting such planets but characterizing them sufficiently that they can be used in a statistical analysis. In particular, the signals from all three categories are all prone to light curve degeneracies that may lead to ambiguities in the planet mass-ratio $q$, separation $s$, and the size of the source $\rho$ (used to measure $\theta_{\rm E}$ and constrain the host mass). Bound planets may also have light curves that are degenerate with models that include a second source rather than a planet. 

The most immediate need for designing the \WF\, Galactic  Bulge Time Domain Survey is a detailed simulation of wide-orbit and small planetary perturbations to investigate how well the planet perturbations will be characterized. This simulation will need to answer the following questions:
\begin{itemize}
	\item{What fraction of the planet detections will be due to central caustic vs. planetary caustic perturbations?}
	\item{What fraction of those will be due to caustic crossings rather than weak distortions?}
	\item{For what fraction of planet perturbations can potentially degenerate light curve solutions be resolved in order to conclusively detect the planet and measure $q$ and $s$ within some tolerance?}
	\item{What cadence of observations is needed to resolve these degeneracies?}
	\item{What fraction of planet detections will be unambiguously characterized?}
\end{itemize}
If there is a large fraction of cases with unresolved degeneracies, the survey area may need to be increased to account for the reduced number of ``useable" planets. This will need to be balanced with the minimum cadence for distinguishing between solutions. In addition, it should be studied whether a mix of cadences over a larger area (as done in ground-based surveys) would be better at balancing these two considerations.

For wide-orbit planets, simulations also need to account for the fact that the stellar microlensing perturbation may be very weak and/or occur outside of the RGBS observing season. Such simulations need to address these questions:
\begin{itemize}
	\item{For what fraction of wide-orbit planets will the microlensing event be detected on the basis of the planetary, rather than the stellar, signal?}
	\item{In cases for which only part of the stellar light curve is captured by RGBS, what are the resulting uncertainties in $t_{\rm E}$ and $\pi_{\rm E}$, and how do those uncertainties affect our ability to measure $q$ and $m_{\rm}$?}
	\item{What cadence of observations between RGBS observing windows is needed to reduce those uncertainties below some tolerance?}
\end{itemize}
It is likely that a cadence of order 1 observation per day or every few days between RGBS observing seasons will be sufficient for characterizing the stellar events. However, unless the pointing tolerances of \WF\, are significantly relaxed, these observations will need to be taken a different observatory. Choosing less extincted fields would enable these observations to be taken from the ground, e.g., with LSST.

These investigations and related trade-studies must be done in order to maximize \WF's ability to take advantage of new parameter space. In particular,
one of the unique features of \WF\, is that the majority of \WF\, planets will come with measurements of the host masses and system distances. This highlights the importance of fully understanding these issues and optimizing the survey to overcome them, because if the degeneracies can be resolved, this will result in direct measurements of the masses and projected semi-major axes of these planets.


\begin{thebibliography}{}
\expandafter\ifx\csname natexlab\endcsname\relax\def\natexlab#1{#1}\fi

\bibitem[{{Abe} {et~al.}(2013){Abe}, {Airey}, {Barnard}, {Baudry}, {Botzler},
  {Douchin}, {Freeman}, {Larsen}, {Niemiec}, {Perrott}, {Philpott},
  {Rattenbury}, \& {Yock}}]{Abe13}
{Abe}, F., {Airey}, C., {Barnard}, E., {et~al.} 2013, \mnras, 431, 2975

\bibitem[{{Gaudi}(1998)}]{Gaudi98}
{Gaudi}, B.~S. 1998, \apj, 506, 533

\bibitem[{{Gaudi} \& {Gould}(1997)}]{GaudiGould97}
{Gaudi}, B.~S., \& {Gould}, A. 1997, \apj, 477, 152

\bibitem[{{Gould} {et~al.}(2023{\natexlab{a}}){Gould}, {Shvartzvald}, {Zhang},
  {Yee}, {Calchi Novati}, {Zang}, \& {Ofek}}]{Gould23_ob1195}
{Gould}, A., {Shvartzvald}, Y., {Zhang}, J., {et~al.} 2023{\natexlab{a}}, arXiv
  e-prints, arXiv:2303.08876

\bibitem[{{Gould} {et~al.}(2023{\natexlab{b}}){Gould}, {Ryu}, {Yee}, {Albrow},
  {Chung}, {Han}, {Hwang}, {Jung}, {Shin}, {Shvartzvald}, {Yang}, {Zang},
  {Cha}, {Kim}, {Kim}, {Lee}, {Lee}, {Lee}, {Park}, \&
  {Pogge}}]{Gould23_kb2397}
{Gould}, A., {Ryu}, Y.-H., {Yee}, J.~C., {et~al.} 2023{\natexlab{b}}, arXiv
  e-prints, arXiv:2306.04870

\bibitem[{{Griest} \& {Safizadeh}(1998)}]{GriestSafizadeh98}
{Griest}, K., \& {Safizadeh}, N. 1998, \apj, 500, 37

\bibitem[{{Hwang} {et~al.}(2018){Hwang}, {Udalski}, {Shvartzvald}, {Ryu},
  {Albrow}, {Chung}, {Gould}, {Han}, {Jung}, {Shin}, {Yee}, {Zhu}, {Cha},
  {Kim}, {Kim}, {Kim}, {Lee}, {Lee}, {Lee}, {Park}, {Pogge}, {KMTNet
  Collaboration}, {Skowron}, {Mr{\'o}z}, {Poleski}, {Koz{\l}owski},
  {Soszy{\'n}ski}, {Pietrukowicz}, {Szyma{\'n}ski}, {Ulaczyk}, {Pawlak}, {OGLE
  Collaboration}, {Bryden}, {Beichman}, {Calchi Novati}, {Gaudi}, {Henderson},
  {Jacklin}, {Penny}, \& {UKIRT Microlensing Team}}]{Hwang18_0173}
{Hwang}, K.-H., {Udalski}, A., {Shvartzvald}, Y., {et~al.} 2018, \aj, 155, 20

\bibitem[{{Jung} {et~al.}(2023){Jung}, {Zang}, {Wang}, {Han}, {Gould},
  {Udalski}, {Albrow}, {Chung}, {Hwang}, {Ryu}, {Shin}, {Shvartzvald}, {Yang},
  {Yee}, {Cha}, {Kim}, {Kim}, {Lee}, {Lee}, {Lee}, {Park}, {Pogge}, {KMTNet
  Collaboration}, {Szyma{\'n}ski}, {Skowron}, {Poleski}, {Soszy{\'n}ski},
  {Pietrukowicz}, {Koz{\l}owski}, {Ulaczyk}, {Rybicki}, {Iwanek}, {Wrona},
  {OGLE Collaboration}, {Green}, {Hennerley}, {Marmont}, {Mao}, {Maoz},
  {McCormick}, {Natusch}, {Penny}, {Porritt}, {Zhu}, {Tsinghua Team}, \& {FUN
  Follow-Up Team}}]{Jung23AF8}
{Jung}, Y.~K., {Zang}, W., {Wang}, H., {et~al.} 2023, \aj, 165, 226

\bibitem[{{Penny} {et~al.}(2019){Penny}, {Gaudi}, {Kerins}, {Rattenbury},
  {Mao}, {Robin}, \& {Calchi Novati}}]{Penny19}
{Penny}, M.~T., {Gaudi}, B.~S., {Kerins}, E., {et~al.} 2019, \apjs, 241, 3

\bibitem[{{Ryu} {et~al.}(2022){Ryu}, {Kil Jung}, {Yang}, {Gould}, {Albrow},
  {Chung}, {Han}, {Hwang}, {Shin}, {Shvartzvald}, {Yee}, {Zang}, {Cha}, {Kim},
  {Kim}, {Lee}, {Lee}, {Lee}, {Park}, \& {Pogge}}]{Ryu22_MP1}
{Ryu}, Y.-H., {Kil Jung}, Y., {Yang}, H., {et~al.} 2022, \aj, 164, 180

\bibitem[{{Suzuki} {et~al.}(2016){Suzuki}, {Bennett}, {Sumi}, {Bond}, {Rogers},
  {Abe}, {Asakura}, {Bhattacharya}, {Donachie}, {Freeman}, {Fukui}, {Hirao},
  {Itow}, {Koshimoto}, {Li}, {Ling}, {Masuda}, {Matsubara}, {Muraki},
  {Nagakane}, {Onishi}, {Oyokawa}, {Rattenbury}, {Saito}, {Sharan}, {Shibai},
  {Sullivan}, {Tristram}, {Yonehara}, \& {MOA Collaboration}}]{Suzuki16}
{Suzuki}, D., {Bennett}, D.~P., {Sumi}, T., {et~al.} 2016, \apj, 833, 145

\bibitem[{{Yee} {et~al.}(2021){Yee}, {Zang}, {Udalski}, {Ryu}, {Green},
  {Hennerley}, {Marmont}, {Sumi}, {Mao}, {Gromadzki}, {Mr{\'o}z}, {Skowron},
  {Poleski}, {Szyma{\'n}ski}, {Soszy{\'n}ski}, {Pietrukowicz}, {Koz{\l}owski},
  {Ulaczyk}, {Rybicki}, {Iwanek}, {Wrona}, {Albrow}, {Chung}, {Gould}, {Han},
  {Hwang}, {Jung}, {Kim}, {Shin}, {Shvartzvald}, {Cha}, {Kim}, {Kim}, {Lee},
  {Lee}, {Lee}, {Park}, {Pogge}, {Bachelet}, {Christie}, {Hundertmark}, {Maoz},
  {McCormick}, {Natusch}, {Penny}, {Street}, {Tsapras}, {Beichman}, {Bryden},
  {Calchi Novati}, {Carey}, {Gaudi}, {Henderson}, {Johnson}, {Zhu}, {Bond},
  {Abe}, {Barry}, {Bennett}, {Bhattacharya}, {Donachie}, {Fujii}, {Fukui},
  {Hirao}, {Ishitani Silva}, {Itow}, {Kirikawa}, {Kondo}, {Koshimoto}, {Li},
  {Matsubara}, {Muraki}, {Miyazaki}, {Olmschenk}, {Ranc}, {Rattenbury},
  {Satoh}, {Shoji}, {Suzuki}, {Tanaka}, {Tristram}, {Yamawaki}, \&
  {Yonehara}}]{Yee21_ob0960}
{Yee}, J.~C., {Zang}, W., {Udalski}, A., {et~al.} 2021, arXiv e-prints,
  arXiv:2101.04696

\bibitem[{{Zang} {et~al.}(2023){Zang}, {Jung}, {Yang}, {Zhang}, {Udalski},
  {Yee}, {Gould}, {Mao}, {Albrow}, {Chung}, {Han}, {Hwang}, {Ryu}, {Shin},
  {Shvartzvald}, {Cha}, {Kim}, {Kim}, {Kim}, {Lee}, {Lee}, {Lee}, {Park},
  {Pogge}, {KMTNet Collaboration}, {Mr{\'o}z}, {Skowron}, {Poleski},
  {Szyma{\'n}ski}, {Soszy{\'n}ski}, {Pietrukowicz}, {Koz{\l}owski}, {Ulaczyk},
  {Rybicki}, {Iwanek}, {Wrona}, {Gromadzki}, {OGLE Collaboration}, {Wang},
  {Zhang}, {Zhu}, \& {MAP Collaboration}}]{Zang23AF7}
{Zang}, W., {Jung}, Y.~K., {Yang}, H., {et~al.} 2023, \aj, 165, 103

\bibitem[{{Zhu} {et~al.}(2014){Zhu}, {Gould}, {Penny}, {Mao}, \&
  {Gendron}}]{Zhu14}
{Zhu}, W., {Gould}, A., {Penny}, M., {Mao}, S., \& {Gendron}, R. 2014, \apj,
  794, 53

\end{thebibliography}
\end{document}